\documentclass[times]{GonnellaLab-StyleArXiv}

\usepackage{blindtext}
\usepackage[overload]{textcase}
\usepackage{amsmath,amssymb}
\usepackage{lmodern}
\usepackage{iftex}
\usepackage{enumitem}
\usepackage{pifont}
\usepackage{booktabs}
\usepackage{multicol}
\usepackage[T1]{fontenc}
\usepackage[utf8]{inputenc}
\usepackage{textcomp} % provide euro and other symbols
\usepackage{color}
\usepackage{fancyvrb}
\usepackage{xcolor}
\usepackage{natbib}
\usepackage{multirow}

\IfFileExists{upquote.sty}{\usepackage{upquote}}{}
\IfFileExists{microtype.sty}{% use microtype if available
  \usepackage[]{microtype}
  \UseMicrotypeSet[protrusion]{basicmath} % disable protrusion for tt fonts
}{}
\makeatletter
\@ifundefined{KOMAClassName}{% if non-KOMA class
  \IfFileExists{parskip.sty}{%
    \usepackage{parskip}
  }{% else
    \setlength{\parindent}{0pt}
    \setlength{\parskip}{6pt plus 2pt minus 1pt}}
}{% if KOMA class
  \KOMAoptions{parskip=half}}
\makeatother
\IfFileExists{xurl.sty}{\usepackage{xurl}}{} % add URL line breaks if available
\IfFileExists{bookmark.sty}{\usepackage{bookmark}}{}
\hypersetup{
  hidelinks,
  pdfcreator={LaTeX via pandoc}}
\DefineVerbatimEnvironment{Highlighting}{Verbatim}{commandchars=\\\{\}}
\ifLuaTeX
  \usepackage{selnolig}  % disable illegal ligatures
\fi

\begin{document}

% Please give the surname of the lead author for the running footer
\leadauthor{Gonnella}

\title{Unambiguosly expressing expectations about the content of prokaryotic genomes}

\shorttitle{Unambiguous expectations about prokaryotic genomes}

% Use letters for affiliations, numbers to show equal authorship (if applicable) and to indicate the corresponding author
\author[1,2\space \Letter]{Giorgio Gonnella}

\affil[1]{Center for Bioinformatics (ZBH), Universität Hamburg, Bundesstrasse 43, 20146 Hamburg}
\affil[2]{Institute for Microbiology and Genetics, Georg-August-Universität Göttingen, Goldschmidtstr. 1, 37077 Göttingen}

\maketitle

%TC:break Abstract
\begin{abstract}
%Lets try to keep the abstract between 70-150 words, I have noticed no guidan
In recent years, the sequencing, assembling and annotation of prokaryotic
genomes has become increasingly easy and cheap. Thus it becomes increasingly feasible and interesting to perform comparative genomics
analyses of new genomes to those of related organisms.
Thereby related organisms can be defined by different criteria, such as
taxonomy or phenotype.

Expectations regarding the contents of genomes
are often expressed in scientific articles describing group of organisms. 
Evaluating such expectations, when a new genome becomes available,
requires analysing the text snippets
which express such expectations, extracting the logical
elements of the text and enabling a formal expression, more suitable
for further automated analyses.

Hereby we present a theoretical framework, alongside practical consideration
for expressing expectations about the content of genomes, with the purpose of enabling such comparative genomics analyses.
The components of the framework include a system for the definition of groups of organisms, supported by a Prokaryotic Group Types Ontology,
a system for the definition of genomic contents, supported by
a Prokaryotic Genomic Contents Definition Ontology.
Finally we discuss how the combination of these two systems may
enable an unambiguous definition of absolute and relative genome content expectation rules.
\end {abstract}
%TC:break main
%the command above serves to have a word count for the abstract


\begin{corrauthor}
    giorgio.gonnella\at uni-goettingen.de
\end{corrauthor}

\begin{multicols}{2}
In recent years, advancements in sequencing, assembling, and annotating prokaryotic genomes have made it easier and cheaper to generate new genome sequences \citep{Carrio2018APO}. The availability of new genomes of groups of organisms for which other genomes have already been previously sequenced provides an opportunity to perform comparative genomics analyses to better understand their evolutionary relationships and highlight their differences \citep{Koonin2008GenomicsOB}.

A comparison of related genomes often relies on the verification of the researcher expectations of certain genomic attributes, such as the presence or absence of certain genes, or the frequency of certain compositional elements in the genome. An example of such expectation are Benchmarking Universal Single-Copy Orthologs, employed in systems for assessing the genome quality \citep{Simo2015BUSCOAG}.

However, to date, there is no general-purpose standard framework for expressing these expectations in a consistent and automated manner, which can be employed for genome comparisons. The expectations of genomic content are often expressed in scientific articles using different terms and structures. The aim of the present study is to address this challenge by presenting a theoretical framework for expressing expectations about the content of genomes in a structured and formalized manner. The framework includes a system for unambiguously defining groups of organisms, for defining the genomic content, and use such definitions for expressing rules of expectations.

Recent developments in the field of computational biology have emphasized the importance of developing structured and formalized systems for the analysis of biological data. Ontologies are often used to provide a structured representation of biological concepts and relationships, which has greatly improved the efficiency and accuracy of analyses in various fields \citep{Smith2007TheOF}. The framework presented in this paper employs ontologies for the definition of organism groups and genomic contents and thus aims at developing a similar structured and formalized system for formal, unambiguous definitions of expectation rules in comparative genomics, which will enable more consistent and efficient comparative analyses in the future.

In particular, first, a system for the definition of groups of prokaryotes is presented, based on a Prokaryotic Group Types Ontology (PGTO). Then, we discuss a method for defining genomic contents. Thereby, we introduce generic definitions, such as \textit{genomic content units} (e.g. a given gene cluster) as objects of measurement (observation or prediction) and \textit{genomic attributes} as variables, for which all aspects for a concrete measurement are defined (e.g. the presence of a given gene cluster on a given plasmid). We thereby introduce a second ontology, named Prokaryotic Genomics Contents Definition Ontology (PGCDO), for supporting the formalization of the genome content description. Finally, we describe how these definitions can be used for the expression of expectation rules.

\section{Defining groups of prokaryotes}

Expectations about the content of genomes always refer to some group of
organisms, whose genomes are the object of the expectation.
An example of such groups are phyla, classes or taxonomic groups at any other rank level. Another example are groups of organisms sharing a common phenotype, such as their trophic strategy or reactivity to Gram stain.

This section describes the information necessary for defining
a group of organisms, in a way suitable for enabling
an operative evaluation of the rules of expectation.

\subsection{Prokaryotic Group Types Ontology}

The way to define groups of organisms depend on the kind of criterion which
defines the group. For instance a taxonomic group could be defined
by linking a taxonomy database, while for a group of organisms
sharing a phenotype, that phenotype could be described.

Since different types of groups exists, it is important to characterize them
and define their relationships. For this purpose, an ontology has been developed, which describes the types
of groups, and the requirement in instances of group definitions
ot these types. It has been named PGTO, which stands for
\textit{Prokaryotic Group Types Ontology}.

For example, one of the nodes in the ontology is 
the group type \textit{taxonomic}. Groups of this type have
as a requirement a link to a
taxonomy database - whenever possible to the NCBI taxonomy database
\citep{Schoch2020NCBITA}.

\subsubsection{Structure and contents of the PGTO}

\begin{table*}[!t]
\centering
\begin{tabular}{p{3cm}lp{10cm}}
\toprule
\textbf{Category} & \textbf{Type} & \textbf{Description / Examples}\\
\midrule
\multirow{6}{*}{\textit{Habitat}} & nutrients level & e.g. oligotrophic \\
& specific nutrients & e.g. sulfur \\
& O2 requirement & e.g. anaerobic \\
& salinity & e.g.\ alophiles \\
& pH range & e.g.\ acidophiles \\
& temperature & e.g.\ thermophyles \\
& kind of habitat & e.g.\ ocean \\
\midrule
\multirow{7}{*}{\textit{Phenotype}} & biological interaction & interaction with other given groups \\
& metabolic & specific metabolic trait (e.g. enzymatic capability) \\
& taxis & tendency to move towards or away from a stimulus \\
& trophic strategy & strategy for obtaining energy and/or organic compounds \\
& resultive disease & causative of a given disease \\
& gram stain & reaction to gram stain \\
& cultiviability & if and how can be cultivated in lab \\
\midrule
\multirow{3}{*}{\textit{Location}} & geographical & e.g. living in the Atlantic Ocean \\
\midrule
\multirow{3}{*}{\textit{Taxonomic}}
& clade & e.g. Proteobacteria \\
& paraphyletic & e.g. Clostridia\\
& strain & e.g. MG1655 \\
& metagenome-assembled & single strain-level taxa from metagenomes\\
\midrule
\multirow{2}{*}{\textit{Derived}} & combined & union or intersection of other groups \\
& inverted & non-members of a given group \\
\bottomrule
\end{tabular}
\caption{Examples of group types in different categories, defined in the Prokaryotic Group Types Ontology (PGTO).}
\label{tab:GroupTypes}
\vspace{5mm}
\end{table*}

The goal of the PGTO ontology are group type definitions. These are summarized in categories, which are children of the root node
\textit{group type categories}.
The categories include:
\begin{itemize}
\item \textit{habitat group type}, i.e. group types by the kind of habitat or required habitat conditions
\item \textit{phenotype group type}, i.e. group types which are defined by some aspect of their phenotype
\item \textit{location group type}, i.e. group types which are defined by the physical location of the organisms
\item \textit{taxonomic group type}, i.e. group types which are defined by an inferred common phylogenetic relation of their members
\item \textit{derived group type}, i.e. group types whose definition depends on another group type definition
\end{itemize}

Examples of group types in the different categories are given in Table \ref{tab:GroupTypes}.

Besides the group categories and group type definitions, the ontology contains metadata terms (namespace \textit{metadata}) and specialized term relationships.

Among the metadata terms are:
\begin{itemize}
\item terms which allow to define
combinations and inversions of group definitions (\textit{groups combination expression}, \textit{group inversion expression} and their ancestor terms \textit{groups logical expression} and \textit{logical expression})
\item terms which allow for definitions to refer to external resources (the generic term \textit{external resource}, the more specific classes \textit{external definition},
\textit{external enumeration},
\textit{external database}, down to specialized terms such as
\textit{taxonomy database} and \textit{strains database})
\item optional attributes of groups, such as \textit{taxonomic rank}
\end{itemize}

The defined relationships are:
\begin{itemize}
\item \textit{defined by}, which is used for describing the way of 
definition of some of the different group types.
\item \textit{has external link to}, with
the optional prefixes \textit{usually}
and \textit{ideally}, which are used for indicating the required or recommended
external links of some group types
(e.g. taxonomy database for a taxonomic group)
\item \textit{has optional type}
indicates that instances of a group may have a value in a given classification, but this is not always the case, e.g. some taxonomic groups have a taxonomic rank (such as species or phylum), some other not, since they are at an intermediate unnamed rank level
\end{itemize}

\subsection{Simple and derived groups}

Often, sentences refer to multiple aspects, which, all together, precisely
define a group of organisms. An examples would be anaerobic Archaea, which
contains two aspects, one is the oxygen requirement (anaerobic)
and one is a taxonomic group (Archaea). This kind of derived groups
are hereby called \textit{combined groups}. The definition of combined
groups can use any kind of set operation, such as intersection or union.

In other cases, a rule can refer to organisms which are not contained
in a given group. In this case, an \textit{inverse group} can be defined as the
group of all organisms outside of that given group. Combined groups definitions
may also use this inversion operation.

Groups which are not derived from combining or inverting other groups
are termed here \textit{simple groups}.

\subsection{Operative considerations}

We consider here, how to operatively define a group of prokaryotes
and give recommendations for the implementation of such
group definitions in software applications.

\subsubsection{Identifiers and names}

In order to be able to unambiguously refer to a group of organisms
when expressing expectations about the content of genomes, the following
elements must be defined.

First an identifier shall be created to allow referring
to the group in different contexts, such as the definition of rules
or of derived groups. This identifier must be unique
among all group definitions.

It is recommended to keep apart from the identifier,
a distinct descriptive name of the group (which should also be unique).
This has the advantage that the identifier must not
necessarily be descriptive
(e.g. it could just be a series of letters and numbers)
and can remain stable,
while the name can change, e.g. when
a taxon is reassigned \citep{Kannan2023CollectionAC}, or a different spelling or format is chosen (e.g. ``Gram +'' instead of ``Gram positive''). It is because
of this kind of consideration, that separate identifiers
and names are kept in existing databases of organisms group, such as NCBI taxonomy \citep{Schoch2020NCBITA}.

\subsubsection{Definition criteria}

Finally, and most importantly, each group must then be exactly defined.
The format of group definition depends on the group type.

For taxonomic groups, the definition can be a reference to a taxonomy 
database. Thereby the NCBI taxonomy can be used, preferentially.
Although this database is not an authoritative source of taxonomic
classification,
it has the advantage, over other more formal taxonomy
sources, to be easy to access. Furthermore,
its use is very convenient,
as it allows to directly find the sequences and annotations in other
NCBI databases, from which genome attributes can be measured.
If a taxonomic group is not defined in NCBI taxonomy, another source can be
used, and its use be documented. For example, for prokaryotic strains,
other databases can be more precise \citep{Reimer2021BacDiveI2}.

For other types of simple groups, links to other external references can be
used. If possible, other sources, listing the species or strains which
belong to the group, shall be used. If not,a description of the group-defining
criteria can be given. Ideally this should be linked to terms in 
existing ontologies, in order to provide a higher precision.

Finally, for derived groups (combined groups and inverse groups),
the logical formula, in terms of set operations,
for deriving that group from other defined groups, should be used as definition.

\subsubsection{Use of the PGTO}

All groups belong to a single group type, which shall be one of leaves
in the subtree under the node \textit{group\_types\_category}
of the PGTO ontology tree.

If no fitting group type exists, then new definitions are required. A group type
definition can be added to the PGTO. The group shall be assigned, whenever possible,
to one of the existing group type categories, and if not, a new category must be
defined.

If new definitions of group types and/or group type categories have been created
(or existing definitions edited), the changes shall be made public,
whenever possible, by sending a pull request to the PGTO Github repository,
in order to avoid clashes with other users of the ontology.

\section{Defining genomic contents}

This study handles about expectations about the content of genomes. Thus it is
necessary to exactly define what is a genomic content, which will be the object
of an expectation.

In order to measure something about the content of a genome,
we have to observe something in its sequence and/or annotation (and eventually
use the observation as base for computations).
Thus multiple aspects can be differentiated logically. The first aspect
of a measurement is the identity of the object of the observation. 
Here we name this object of the observation \textit{genomic content unit}.

Furthermore, for obtaining an observation value, we need, besides
the object of observation, to define more
detail, e.g. the measurement mode, the genomic regions to consider,
and for relative measurements, the reference object to which to relate.
Here we name this well defined aspect of the genomic content, for which a value
can be determined, a \textit{genomic attribute}.

\subsection{Genomic content units}

\begin{table*}[!t]
\centering
\begin{tabular}{p{3cm}lp{10cm}}
\toprule
\textbf{Category} & \textbf{Type} & \textbf{Description / Examples}\\
\midrule
\multirow{4}{*}{\textit{Sequence}}
& DNA base & any base, or all G and C \\
& $k$-mer & a given sequence of length $k$ \\
& molecule & any molecule, any plasmid, or a given plasmid \\
& alignment region & sequence region similar to a given sequence \\
\midrule
\multirow{5}{*}{\textit{Single feature}}
& protein-coding gene & ammonia monooxygenase (\textit{amoA}) \\
& RNA gene & 16S rRNA gene \\
& operon & plipastatin operon (\textit{ppsABCDE}) \\
& genomic island & cytotoxin-associated gene pathogenicity island (\textit{cag} PAI) \\
\midrule
\multirow{6}{*}{\textit{Feature category}}
& feature type & pseudogene (SO:0000336) \\
& protein domain & diguanylate cyclase (GGDEF, IPR000160) \\
& enzymatic activity & hydrogenase (EC 1.18.3.1) \\
& orthology group & arsenical permease (COG1055) \\
& functional category & signal transduction mechanisms (COG category T) \\
& gene/protein family & PPE gene family (PF00823) \\
\midrule
\multirow{2}{*}{\textit{Feature set}}
& metabolic pathway & MEP/DOXP pathway (KEGG map00900)\\
& gene cluster & sulfor oxidation gene cluster (sox) \\
& feature arrangement & 16S rRNA, tRNAs, 23S rRNA\\
\bottomrule
\end{tabular}
\caption{Examples of different types of genomic content units,
defined in the Prokaryotic Genomic Contents Definition Ontology (PGCDO)}
\label{tab:UnitTypes}
\vspace{5mm}
\end{table*}

\textit{Genomic content units} are defined in this study as all possible objects
of observation, contained in, or derivable from the genome sequence and/or 
the genome annotation.

They can have different degrees of complexity, as in the following examples:
any base of the DNA sequence; a specific k-mer; a gene; all features of
a given type; an enzymatic activity derived from the product of multiple genes;
a metabolic pathway involving different enzymes; the relative
arrangements of some given features in the genome sequence.

A classification of different types of genomic content units is presented in
the next sections. 
Examples of the different types of units are given in Table
\ref{tab:UnitTypes}.

\subsection{Sequence-based genomic content units}

Genomic content units which involve only the sequence are the following:
\begin{itemize}
\item single DNA bases (eventually of a given type, or of one of a set of different types)
\item adjacent sequences of a given length ($k$-mers)
\item molecules
\item parts of the sequence fulfilling some criteria (e.g. matching a given pattern)
\end{itemize}

These units are employed in the definition of basic sequence statistics, such as the number of molecules or
the total sequence length, and compositional sequence statistics, such as the
$k$-mer spectrum or the GC content.

\subsection{Annotation-based genomic content units}

Annotation-based genomic content units are those units which involve
one or multiple features of the genomic annotation.
Thereby we consider both features which are structural parts of the
sequence (i.e. regions or combinations of regions
of the genome) and their products (e.g. the protein coded by a gene and its
function).

From a quantitative point of view, one can distinguish between
simple features,
categories of features which are equivalent according to some
 aspect (structural or functional) and sets of features which
contain multiple members, which are not equivalent, but rather express
distinct subunits, either with some criteria regarding their relative
positioning (ordered feature sets) or not (unordered feature sets).

\subsubsection{Feature categories}

\textit{Feature categories} are here defined as multiple features
which are considered by a single aspect in which they are equivalent to each other.
They share a single category-defining characteristic,
such as structure or function.

Examples of feature categories are
all feature belonging to a given \textit{sequence feature type},
to a given \textit{ortholog group}, or whose product has
a given \textit{enzymatic activity}.

\subsubsection{Feature sets}

In one type of feature collection, the identity of the single
components is the only criteria to take into consideration in order to characterize
the genomic content unit.
In this case we speak of \textit{unordered feature sets}.

An example of feature set is a metabolic pathway. For example, according to \citet{Pasternak2013ByTG}, enzymes forming the DOXP pathway are absent
from most predatory bacteria. To verify this rule, one would require to list the genes which code for the enzymes of the pathway and identify
them in the genomes.

In other cases, besides the identity of the single components,
also the relative arrangement of the components to each other is necessary.
An examples are gene clusters, in which genes with a related function are close to
each other. In this case we speak of \textit{ordered feature sets}. 

For example \citet{Giraud2007LegumesSA} explains that the photosynthesis gene cluster (PGC), is common in purple photosynthetic bacteria. To express this rule formally, the required and optional members of the gene cluster in different organisms should be listed. The proximity requirement in this case is that the genes are all close to each other and no other, or few other genes are in between.

Also, one can describe in this way an expectation of gene arrangements. For example, in \citet{McLeod2004CompleteGS} is stated that in Rickettsia, the 16S rRNA gene was separated from the 23S and 5S rRNA genes. This can be translated in the set with the 3 elements, with the relative location constraint of proximity for 23S to 5S and of separation for 23S to 16S.

\subsubsection{Operative considerations}

Regarding simple features, the following consideration must be done.
Since the purpose of this system is to define rules of expectations which are
used to compare the contents of different genomes, it can be difficult to precisely
define a simple feature.
For example, in different genomes, different names can be used for
the same gene. Thus, usually a mapping of rules for single features
to feature categories, such as ortholog groups, or protein families
is required.

An exact operative definition of a category features
involves the description of the category-defining criterion. In general, this
can refer to an external
source, e.g. the name of an ortholog group, or the name of a sequence
feature type.

When defining feature sets, the composition in terms of elements which belong
to the set must be given. Not all elements must be present necessarily all the times.
The definition can sometimes involve complex rules, such as the presence of
one possible element, excluding another.
For ordered sets, besides the composition, also the relative location of the elements,
i.e.\ the feature relative order, and their proximity requirements, must be given.

\subsection{Defining genomic attributes}

When considering the contents of the genome in different genomes, we need to define
multiple aspects of the measurement, in order to produce a value which can
be compared. We name here the object of a measurement, for which a value
can be given, and which involves one or more genomic content units,
a \textit{genomic attribute}.

An exact definition of a genomic attribute must always include the following elements:
\begin{itemize}
\item an \textit{measurement object} is the genomic content unit that is observed and measured
\item the \textit{measurement mode}, for example absolute count, relative frequency, presence/absence, total sequence length,
  conservation, completeness
\item the \textit{measurement region} where the genomic unit is considered, which can be the entire genome, a type of molecule (chromosome, plasmid), a specific molecule (e.g. a specific plasmid), a feature type (e.g. coding regions) or the surroundings of a given feature
\end{itemize}

Depending on the measurement mode, some more elements must be included:
\begin{itemize}
\item the \textit{measurement reference} for relative measurements (e.g. frequency, conservation) the genomic content units to which the measurement is relative
\item the \textit{subunit definition} for measurements modes which
involve counting a subunit (e.g. exons in coding sequences)
\end{itemize}.

\subsection{Prokaryotic Genomic Contents Definition Ontology}

In order to provide a stable framework for the definition of genomic contents, using the system described above, a Prokaryotic Genomic Contents Definition Ontology (PGCDO) has been created.

Besides the basic definitions given above (genomic attribute, genomic region, measurement mode, genomic content unit), the ontology contains
three sections. Nodes belonging to the namespace
\texttt{genomic\_region\_type} define the different types of genomic regions,
such as the whole genome, genomic regions defined by specific molecule
names or types. In the namespace \texttt{measurement\_mode} are the definitions
of different types of measurement mode, i.e. absolute count, relative frequency, presence/absence, total or average sequence length,
and number of subunits (e.g. number of exons in some kind of genes).

Finally under the node \texttt{genomic\_content\_unit} are the
categories of genomic content unit types described above (sequence and annotation based, the latter consisting of single features or different types of feature collections). Genomic content unit types are then
given as terms in the namespace
\texttt{genomic\_content\_unit\_type}.
Not all possible genomic content unit types
are given, in particular not those under the node \texttt{feature},
as these are rather better
defined by a link to a term of the Sequence Ontology \citep{Eilbeck2005TheSO}.

\section{Defining rules of expectation}

The goal of the system presented in this study is to represent rules of expectations about the
genomic contents of prokaryotes.

Given the systems described above for unambiguosly define groups of organisms and
genomic, different kind of expectation rules can now be formulated.
A distinction must be done, regarding to what the value of the considered attribute must be
compared.

\subsection{Absolute expectations}

In \textit{absolute expectations} the value is compared to a reference.
This can be a single reference value, a set of values or value range.
Furthermore, an expected quantitative relation is given.
Examples of such relations are: larger, smaller, equal (with one reference value),
one of (with a set of reference values), included in, excluded from
(with a range of reference values).

For example in \citet{McLeod2004CompleteGS} it is stated that members of the Rickettsia taxon have comparatively small genomes (1.1 to 1.3 Mb). Although the word comparatively is used, the comparison term is kept implicit and instead, a range of expected values is given.

\subsection{Comparative expectations}

In \textit{comparative expectations} the expectation
of the value of an attribute is not expressed by giving the expected
relation of the values in two different groups of organisms.
Also in this case an expected quantitative relation is given. Not all relation
operators used for absolute expectations are useful here (e.g. one of). Instead
only relations applying to a single value (if the reference group has always
a given value) or a value range (the values for the group) can be used.

For example an analysis \citet{Lauro2009TheGB} showed that the genomes of  copiotrophic marine bacteria contain more
repeats within clustered regularly interspaced short palindromic repeats (CRISPRs) than that of oligotrophs.

\subsection{Group portion}

For both relative and absolute expectations, the strength of the expectation must be given in terms of which portion of the group meets the expectation. For example an expectation can apply to all,
most or some of the organisms of a group. In some cases a concrete value, say 30\%, is given.

Although expectations which reflect only a portions of the organisms of a
group are less informative, they can still be useful.
For example in \citet{Hahn2012ThePY} is stated that most genomes of \textit{Cupriavidus} and \textit{Ralstonia} strains contain genes for
putative carbohydrates transporters. If the genome of a new strain
in these groups do not include such transporters, this information
will not be completely surprising but can still be interesting.
If many other such negative examples were found, for example, than
one could conclude that the stated rule would not held anymore.

\section*{Discussion and Conclusions}

The framework proposed in this paper creates a formal system for expressing expectations about the content of genomes, as a base for automated and standardized comparative genomic analyses. The goal
of this framework is to represent a foundation for future comparative genomics studies, enabling researchers to more efficiently and effectively compare the genomes of different organisms, by defining rules of
expectation about their contents and enabling the evaluation
of such expectations, which, in turn, can highlight unusual
and unexpected characteristics of a genome.

In particular, the framework consists of two base components, a system for the definition of organism groups and one for the definition of
genomic contents in form of measurable attributes. These components
are then combined in the form of rules
of expectations, putting in relation the expected attribute value in
a group of organisms to a given reference, in form of concrete values
or comparison to another group of organisms.

The system developed in this paper was specifically designed for the analysis of prokaryotic genomes, reflecting the focus of the project it was developed in. However, the underlying principles and the framework established could in theory be extended to the analysis of eukaryotic genomes as well. This could be a future direction for the development of the system, and will require further investigations into the unique challenges and requirements of eukaryotic genomes.

In addition to the definition of the logical elements of the genome expectations as presented in the paper, a concrete use of the framework
in the context of computational biology will require to consider
further aspects. In particular, a practical implementation 
will require a definition of the data model, which defines how the data, in this case the genome expectations and their components, is structured and stored, for example in a database and by a file format. This structure must be able to effectively represent the relationships between the different elements, such as the connection between a group of organisms and the expectations rules, or the relationship between a genomic content unit and the corresponding genomic attribute. A well-designed data model allows for efficient querying and manipulation of the data, and ensures the reliability and consistency of the stored information.

In the data model it will also be necessary to consider further metadata, not handled in the theoretical framework presented in this paper. In particular, it is also important to consider the need for documenting the sources of the expectations, such as snippets from scientific literature, records in biological database or even other possible sources, such as expert knowledge. It is crucial to have a robust and transparent way of tracking the sources of the expectations to ensure the credibility and reproducibility of the results. An effective implementation system must carefully balance the technical and organizational requirements of the data model and source tracking with the flexibility and expressiveness of the framework for expressing expectations.

Finally, the use of the expectation rules will require providing clear operating instructions on how to perform the computations of genome attribute values, for example programming code for each attribute, and the development of a batch computation system and an appropriate attribute values storage. 

\bibliography{references}

\begin{avail}
The Prokaryotic Group Types Ontology is available in the GitHub repository at the URL
\url{https://github.com/ggonnella/pgto} in OBO format.
The Prokaryotic Genomic Contents Definition Types Ontology is available in the GitHub repository at the URL
\url{https://github.com/ggonnella/pgcdo} in OBO format. Both are released under the open source ISC license.
\end{avail}

\begin{acknowledgements}
Giorgio Gonnella has been supported by the DFG Grant GO 3192/1-1 ‘`Automated characterization of microbial genomes and metagenomes by collection and verification of association rules’’. The funders had no role in study design, data collection and analysis, decision to publish, or preparation of the manuscript.

Many thanks to Serena Lam for helping identifying examples of expectation rules in the scientific literature which were added to the manuscript.
\end{acknowledgements}

\begin{contributions}
 These contributions follow the Contributor Roles Taxonomy guidelines: \href{https://casrai.org/credit/}{https://casrai.org/credit/}.
 Conceptualization: G.G.;
 Data curation: G.G.;
 Formal analysis:  G.G.;
 Funding acquisition:  G.G.;
 Investigation: G.G.;
 Methodology: G.G.;
 Project administration: G.G.;
 Resources: G.G.;
 Software: G.G.;
 Supervision: G.G.;
 Validation: G.G.;
 Visualization:  G.G.;
 Writing – original draft: G.G.;
 Writing – review \& editing: G.G.
\end{contributions}

\begin{interests}
 The authors declare no competing financial interests.
\end{interests}

\end{multicols}

\end{document}